\documentclass[lettersize,journal]{IEEEtran}
\usepackage[noadjust]{cite}
\usepackage{braket}
\usepackage{mdwmath}
\usepackage{varwidth}
\usepackage{float}
\usepackage{placeins}
\usepackage{afterpage}
\usepackage{ragged2e}

\usepackage{eqparbox}
\usepackage{lipsum}
\newcolumntype{P}[1]{>{\centering\arraybackslash}p{#1}}
\usepackage{epsfig}
\usepackage{epstopdf}
\usepackage{array}
\usepackage{subcaption}
\usepackage{hyphenat}
\usepackage{physics}
\usepackage{cuted}
\usepackage{enumitem}
\usepackage{booktabs} 
\usepackage{bbding}
\usepackage{pifont}
\usepackage{wasysym}

\usepackage{framed,url,caption,graphicx}
\usepackage{amsmath,xcolor,soul,algpseudocode,algorithm,xspace,tabularx,multirow}
\usepackage{amssymb, cleveref}
\usepackage[utf8]{inputenc}
\usepackage[english]{babel}
\usepackage{fancyhdr}
\usepackage{parselines} 
\usepackage{blindtext}
\usepackage{mathtools}

\title{Improving Injection-Throttling Mechanisms for Congestion Control for Data-center and Supercomputer Interconnects}

\author{Cristina Olmedilla\textsuperscript{1}, Jesus Escudero-Sahuquillo\textsuperscript{1}, Pedro J. Garcia\textsuperscript{1}, Francisco J. Quiles\textsuperscript{1} and
Jose Duato\textsuperscript{2}
}

\begin{document}
\maketitle
\begingroup
\renewcommand\thefootnote{\arabic{footnote}}
\footnotetext[1]{Department of Computing Systems, Universidad de Castilla-La Mancha, Spain. Email: cristina.olmedilla@uclm.es}
\footnotetext[2]{Royal Spanish Academy of Sciences, Spain}
\endgroup

\begin{abstract}
Over the past decade, Supercomputers and Data centers have evolved dramatically to cope with the increasing performance requirements of applications and services, such as scientific computing, generative AI, social networks or cloud services. This evolution have led these systems to incorporate high-speed networks using faster links, end nodes using multiple and dedicated accelerators, or a advancements in memory technologies to bridge the memory bottleneck.
The interconnection network is a key element in these systems and it must be thoroughly designed so it is not the bottleneck of the entire system, bearing in mind the countless communication operations that generate current applications and services.
Congestion is serious threat that spoils the interconnection network performance, and its effects are even more dramatic when looking at the traffic dynamics and bottlenecks generated by the communication operations mentioned above.
In this vein, numerous congestion control (CC) techniques have been developed to address congestion negative effects.
One popular example is Data Center Quantized Congestion Notification (DCQCN), which allows congestion detection at network switch buffers, then marking congesting packets and notifying about congestion to the sources, which finally apply injection throttling of those packets contributing to congestion. 
While DCQCN has been widely studied and improved, its main principles for congestion detection, notification and reaction remain largely unchanged, which is an important shortcoming considering congestion dynamics in current high-performance interconnection networks.
In this paper, we revisit the DCQCN closed-loop mechanism and refine its design to leverage a more accurate congestion detection, signaling, and injection throttling, reducing control traffic overhead and avoiding unnecessary throttling of non-congesting flows.

\end{abstract}

\section{Motivation}

Supercomputers and data-centers (DCs) have evolved substantially over the past decade, and they will continue to expand in the near future~\cite{panda2018networking}, as we move from the Exascale era towards Zettascale, characterized by heterogeneous architectures, cloud-native workflows, and the pervasive use of artificial intelligence (AI) across a wide range of applications. Beyond AI, domains such as large language models (LLMs), digital twins, computational fluid dynamics (CFD), climate modeling, defense, physics, and space science are also key workloads for Supercomputers or DCs.

The interconnection network is the backbone of Supercomputers (SCs) and Datacenters (DCs), where thousands of computing or storage endpoints (i.e., CPUs, GPUs, HDDs, or NVMe), cooperate to run the mentioned computing- and data-hungry applications. These applications generate countless communication operations that the network must process as quickly as possible, or the network may become the system's bottleneck. Interconnection networks are composed of interfaces (i.e., NICs) at endpoints generating traffic flows, and switches that interconnect all of them using high-speed
links. These endpoints are connected within heterogeneous nodes via an intra-node network, and hundreds of these nodes are interconnected through the inter-node network.

Historically, Supercomputers and DCs have diverged in their different purposes, but they have shared some characteristics; they were largely centralized, with limited cloud infrastructure; their hardware platforms mainly were homogeneous, with little architectural diversity; and their interconnection networks typically provided bandwidths in the 10--40~Gbps range on DCs and 40--100~Gbps on HPC.
However, in recent years, these infrastructures have evolved to be predominantly cloud-based, with heterogeneous systems, such as edge computing and globally distributed DCs. At the same time, their interconnection networks have increased their link bandwidth to 100--800~Gbps~\cite{Hoefler22convergence}. Furthermore, the widespread adoption of GPU accelerators across both supercomputers and hyperscale DCs has effectively bridged the two ecosystems from the network architecture perspective. For example, AI/LLM training and simulation pipelines rely on collective communication patterns (e.g., all-reduce, all-to-all) that benefit from RDMA over Converged Ethernet (RoCEv2)~\cite{guo2016rdma}.
In this context, the interconnection network must guarantee high bandwidth and low tail latency to communication operations of these applications, while minimizing energy consumption. 

Unfortunately, failures in network devices, aggressive power management, and, most importantly, network congestion are likely to occur as the size of current supercomputers and DCs increases, thereby spoiling the performance of the interconnection network. Specifically, congestion occurs when applications generate communication operations whose inherent traffic flows clog multiple internal network paths, overloading switch buffers. Other causes of congestion include rerouting traffic around faulty regions or lowering link speed to save power. Due to switch buffer clogging, flow-control backpressure propagates congestion from the switch where it originates (i.e., the root) back to the server end nodes generating the congesting flows. Regardless of its cause, congestion affects communication operations that do not directly contribute to network clogging (i.e., victims). This effect is known as head-of-line (HoL) blocking, which impacts network throughput, communication latency, power consumption, and more interconnection network features.

Since the evolution of congestion trees (i.e., congestion dynamics) depends on each application’s traffic pattern, it is challenging to address congestion effects, such as HoL blocking or buffer hogging. Consequently, efficient congestion control (CC) techniques are required to mitigate the impact of congestion and its side effects on network performance.
One of the most widely used strategies to address congestion involves monitoring the occupancy of switch buffers. When the occupancy of a buffer exceeds a given threshold, the packets populating it are considered congestion contributors. The network must notify the source end nodes generating those packets to reduce the injection rate for the traffic flows to which they belong. In this manner, congestion trees are drained from the network, thereby removing the congestion and its effects. This CC approach, known as injection throttling, has been commonly deployed in the last two decades in network technologies such as InfiniBand~\cite{pfister2001introduction} and Ethernet~\cite{rfc3168}.

As network technologies have been improved to overcome interconnection network design challenges, it is natural to expect that the CC mechanisms have evolved accordingly. For instance, the Data-Center Quantized Congestion Notification (DCQCN)~\cite{Zhu15dcqcn} is an injection throttling technique designed explicitly for Ethernet-based networks and protocols (e.g., RoCEv2) and has been widely studied and deployed in current interconnection network technologies. Although vendor-specific DCQCN parameters may vary, their core philosophy persists in modern network technologies. Now it operates in both worlds: cloud AI clusters and mixed HPC installations, where GPU-driven, bursty traffic or shallow-buffer switches make precise, robust congestion control essential.

Specifically, DCQCN comprises mechanisms for congestion detection, signaling, and elimination, known as \emph{Congestion Point} (CP), \emph{Notification Point} (NP), and \emph{Reaction Point} (RP), respectively, which need to be revisited and re-designed to cope with the congestion dynamics of current applications and services. 
In this paper, we present several improvements to these mechanisms. Regarding CP, congestion detection is based on buffer occupancy and does not accurately differentiate between packets that contribute to congestion and those merely affected by it. We propose a more accurate congestion detection mechanism, called Enhanced Congestion Point (ECP)~\cite{olmedilla2024ecp,olmedilla2025ecp}, that eliminates the incorrect marking of victim packets. Regarding NP, note that it is required to ignore the majority of marked packets to avoid flooding the interconnect network with control traffic. Without an accurate congestion-detection mechanism, we may miss important information when aggressively filtering congestion notifications. The use of ECP, combined with our enhanced NP (ENP) mechanism, allows congestion notification packets (CNPs) to be sent only to the source end nodes that are genuinely contributing to congestion.
Regarding RP, DCQCN employs conservative rate adjustments because CNPs lack per-flow context. During congestion, it is highly probable that ``victim'' flows are throttled alongside congesting ones. And, due to an expanded throttled situation, their near-synchronous recovery may recreate congestion cycles or lead to an underutilized network. We propose an enhanced RP (ERP) mechanism that accurately adjusts the injection rate of congesting flows based on more accurate, timely congestion severity information.

We have modeled the improved DCQCN mechanisms in a simulation framework and performed a large set of experiments under realistic workloads. These experiments demonstrate that our proposed ECP, ENP, and ERP mechanisms significantly outperform congestion detection, signaling, and throttling, thereby allowing the injection rate of congesting flows to be adjusted according to available bandwidth, ensuring that the victim flows expected performance is not affected.

\section{Evaluation}\label{sec:result}
In this section, we provide a small scenario so that we can see from the detail how our mechanisms works.

\subsection{Small network experiment}

\begin{figure}[!htb]
\centering
     \begin{subfigure}{.85\columnwidth}
        \centering
        \includegraphics[width=1\columnwidth]{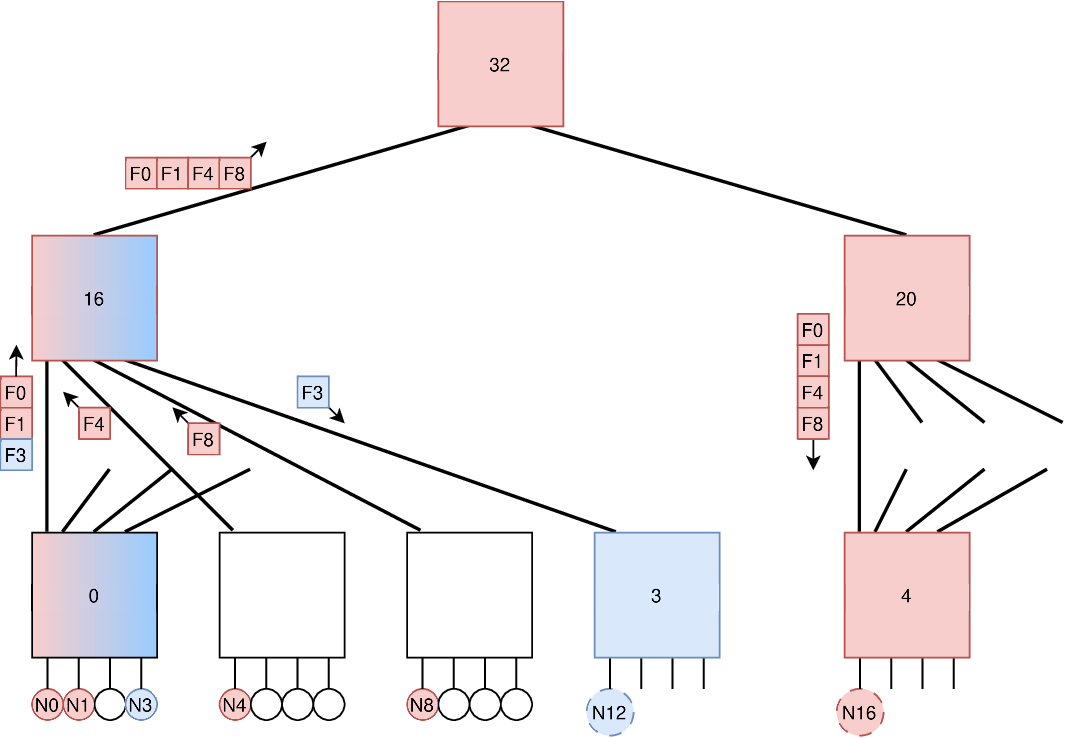}
    \end{subfigure}
    \caption{\small{64 nodes interconnection network, CLOS Topology with 3 stages.}}
    \label{fig:net}
\end{figure}

For this experiment, we provide the network scheme in Fig.~\ref{fig:net}. We have used a serial full-duplex pipelined link of 100 Gbps and 25~ns of link propagation delay, i.e., the time spent by the first cell of each packet in crossing the link. 
For these switches we have modeled 64MB shared buffers and each switch port has at least 512KB of buffer capacity and packet MTU of 1KB\footnote{These values have been obtained based on the parameters of the Broadcom Tomahawk 3~\cite{wheeler2019tomahawk}.}.
We have also modeled a deterministic routing similar to $D$-mod-$K$, which balances the routes of the flows so that the links at a given stage are crossed by a similar number of flow routes. We have modeled a 64-node CLOS topology, composed of 48 switches using 8 ports in 3 stages.

Regarding the DCQCN parameters, we have assumed a single detection threshold, i.e., $K_{min} = K_{max} = V$, with $V=15~KB$. For the rest of the parameters, we have configured the values defined in~\cite{Zhu15dcqcn} in the fluid model.
Regarding the traffic model, as it is illustrated in Figure \ref{fig:net}, we assume four flows F0, F1, F4 and F8 that inject traffic at full speed (i.e., 100~Gbps or 12.5~GB/s) from nodes N0, N1, N4 and N8, respectively, to the destination node N16 generating an incast situation. 
We also assume a "victim" flow F3 that communicates source node N3 with the destination node N12, which is also generated at full speed. All of them are active from $1ms$ (from 2ms to 3ms).
This simple scenario shows a congestion tree that clogs input buffer of switch 16, where packets from congesting flows F0 and F1 generate HoL blocking to packets from flow F3.

\subsection{Results}

\begin{figure}[!htb]
\centering
    \begin{subfigure}{.9\columnwidth}
        \centering
        \includegraphics[width=.9\columnwidth, trim={12cm 0cm 12cm 0cm},clip]{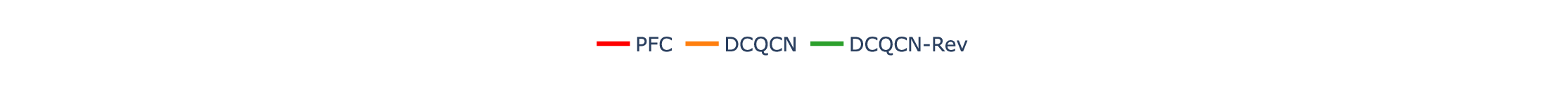}
    \end{subfigure}\\

     \begin{subfigure}{.85\columnwidth}
        \centering
        \includegraphics[width=1\columnwidth]{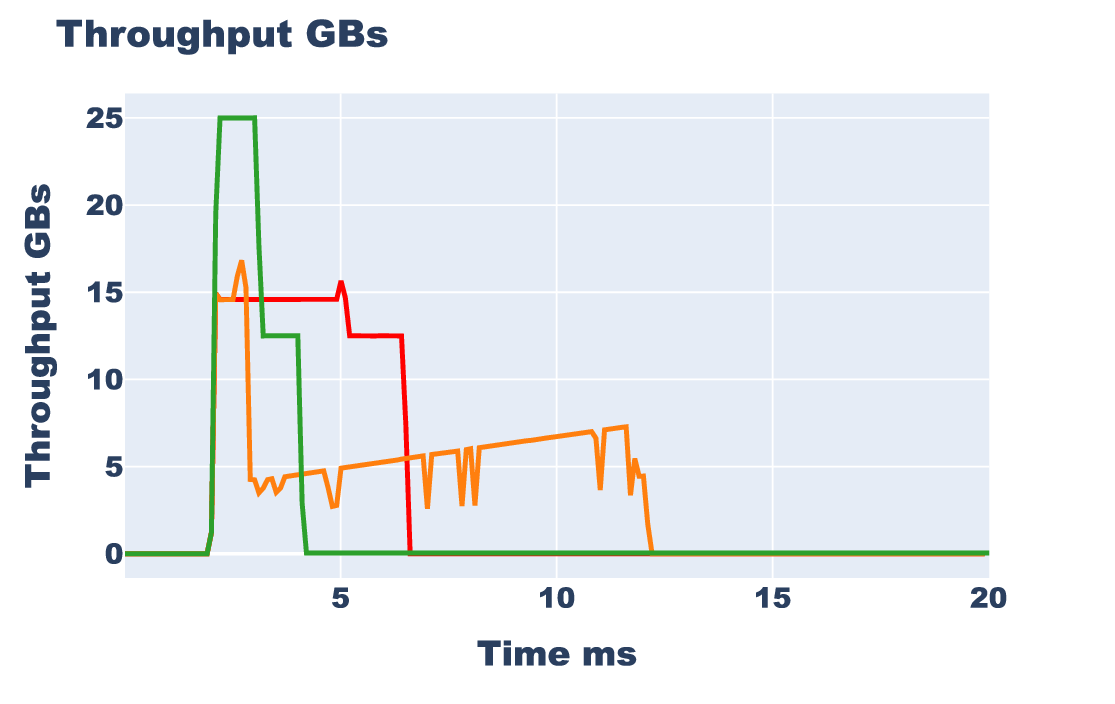}
    \end{subfigure}
    \caption{\small{Network throughput for the 64-nodes scenario.}}
    \label{fig:throughput}
\end{figure}

Fig.~\ref{fig:throughput} shows the network throughput for both congesting and non-congesting flows.
As we can see, our proposal (i.e., DCQCN Rev) is able to maximize network throughput when all the traffic flows are injected in the network. 
First, the maximum network throughput when all the flows are active is 25~GB/s, since flow F3 (i.e., the victim flow) achieves 12.5~GB/s, while all the flows contributing to the incast at node N12 can only receive a maximum of 12.5~GB/s, which should be fairly shared among flows F0, F1, F4 and F8 (i.e., $12.5/4=3.125~GB/s$.
Most importantly, the network is able to deliver all the packets for all these flows before $4$~ms of runtime, while PFC finishes in the $6.5$~ms instant and DCQCN around $12.5$~ms.

Note that PFC, i.e., without congestion management achieves a maximum overall throughput of $15$~GB/s, since the victim flow packets suffer from HoL blocking at switch 16, caused by packets from congesting flows F0 and F1, thus the victim flows performance being significantly reduced.
The same happens with DCQCN, since it is not able to accurately detect, notify and throttle the injection rate of congesting flows.
When congestion appears, victim packets are also marked, and all the traffic flows (congesting and victim) are throttled.
Furthermore, as the congestion tree severity information is not precise enough, congesting flows are throttled too much either in time and bandwidth.

\begin{figure*}[tb]
\centering
    \begin{subfigure}{.9\textwidth}
        \centering
        \includegraphics[width=1\textwidth, trim={4cm 0cm 4cm 0cm},clip]{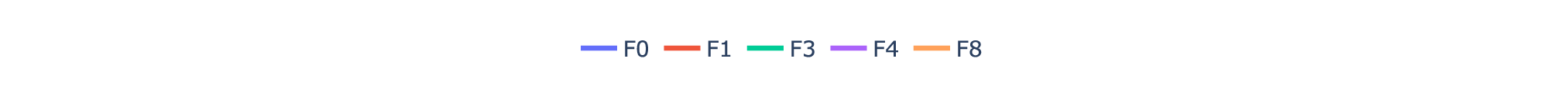}
    \end{subfigure}\\

     \begin{subfigure}{.9\textwidth}
        \centering
        \includegraphics[width=1\textwidth]{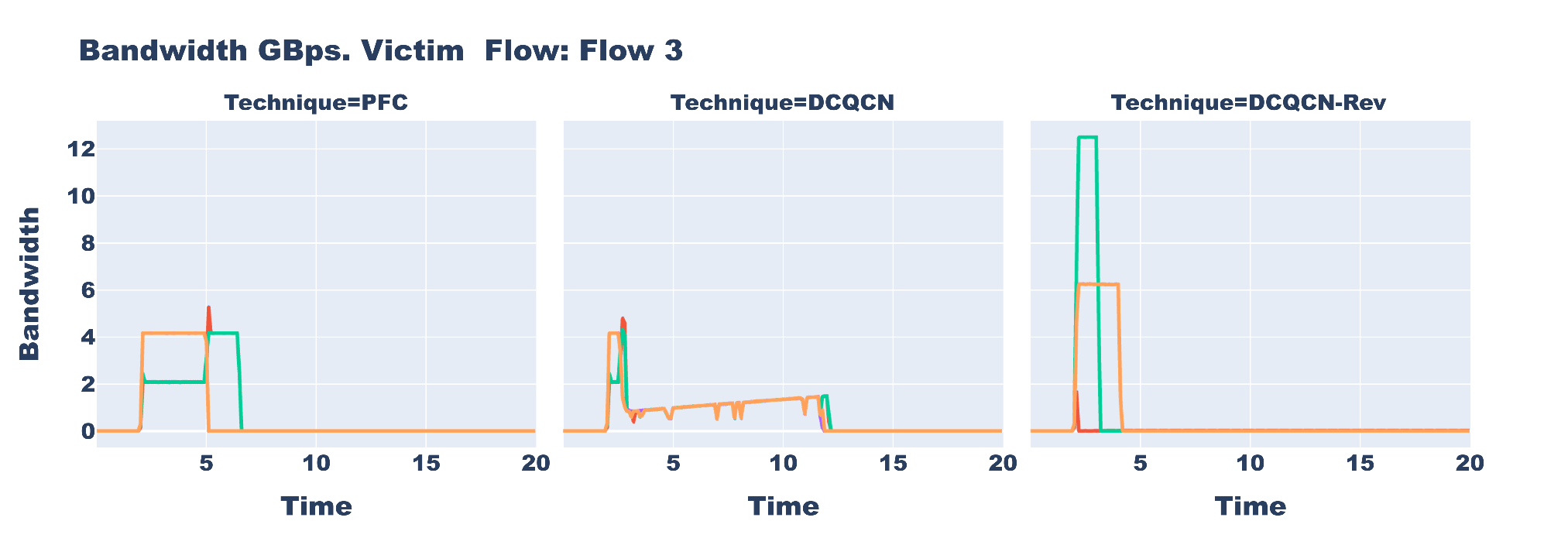}
    \end{subfigure}
    \caption{\small{Network bandwidth per flow for the 64-nodes scenario.}}
    \label{fig:flow}
\end{figure*}

Figure~\ref{fig:flow} illustrates the network bandwidth per traffic flow.
As we can see, victim flow F3 is affected differently across mechanisms. With PFC, the classic parking-lot pathology appears on F0, F1, and F3 (flows that share the same egress port) so, in the absence of end-to-end congestion control, backpressure degrades the performance of the victim flow. DCQCN detects congestion and throttles injection, removing the parking-lot buildup, yet the victim flow is still affected. In addition, we can observe that due to its conservative recovery, this technique leads to prolonged underutilization of the fabric. 
Our proposal DCQCN-Rev achieves the best results for the victim packets, while fairly sharing the remainder network bandwidth among the congesting flows.

\section{Conclusion}

Congestion control (CC) is necessary in high-performance interconnection networks of supercomputers and data centers (DCs) to keep the network performance when communication operations of current applications and services clog certain network parts and spoil the network performance.
Injection throttling, and particularly DCQCN, is a CC approach widely used and deployed by commercial interconnect solutions (such as InfiniBand or Ethernet), which is based on detecting, signaling and throttling congesting flows.
DCQCN defines a three mechanisms: a congestion detection threshold (CP) that measures the switch buffers occupancy so that when the number of stored packets exceeds a given thresholds, the switch marks them as congesting. These marked packets reach the destination, the congestion notification mechanism (NP) sends congestion notification packets (CNPs) back to the source end nodes that generated the marked packets. When those end nodes receive the CNPs, they will reduce the injection rate accordingly based on the reaction point (RP) policy.
DCQCN has numerous drawbacks such as imprecise congesting packet marking, no timely CNPs provision and obsolete reaction at source end nodes.
In this paper, we have proposed an enhanced version for the DCQCN proposal based on an accurate congesting packets marking, timely congestion notification and accurate reaction to only throttle the injection of congesting flows, leaving harmless the congestion effects (i.e., the HoL blocking).

\section*{Acknowledgments}
This research publication is part of the TED2021-130233B-C31 project, funded by MCIN/AEI/10.13039/501100011033 and by the European Union “NextGenerationEU”/PRTR and PERTE-Chip grants (UCLM Chair, TSI-069100-2023-0014) funded by the Spanish Ministry of Digital Transformation and Public Service.

\bibliographystyle{IEEEtran}
\bibliography{bib}
\end{document}